\documentclass[pre,twocolumn,twoside,byrevtex,superscriptaddress,floatfix]{revtex4-1}

\usepackage{currfile}
\usepackage{color}

\usepackage{graphicx,epsfig,verbatim,enumerate}
\usepackage{amssymb,amsmath}
\usepackage{ifthen}

\usepackage{longtable}

\usepackage{mathtools}

\newboolean{twocolswitch}

\newcommand{\tavg}[1]{\langle#1\rangle}

\newcommand{\sindex}[1]{}
\newcommand{\nindex}[1]{}

\newcommand{\www}[1]{\url{#1}}

\newcommand{\Req}[1]{Eq.~(\ref{#1})}

\usepackage{lettrine}

\newcommand{\dee}[1]{\textnormal{d}#1}

\newcommand{\Prob}[1]{{\textnormal{Pr}}\left(#1\right)}

\newcommand{\dstar}{d^\ast}
\newcommand{\dstari}{d_i^\ast}

\newcommand{\phifix}{{\phi^{\ast}}}

\newcommand{\infectionprob}{p}
\newcommand{\infectionprobcrit}{\infectionprob_{\textnormal{c}}}

\newcommand{\distributionfunction}[1]{P^{\textnormal{(#1)}}}

\newcommand{\dosedistribution}[1]{\distributionfunction{dose}_{#1}}

\newcommand{\dosedistributionbare}{\distributionfunction{dose}}

\newcommand{\thresholddistribution}[1]{\distributionfunction{thr}_{#1}}
\newcommand{\thresholddistributionCDF}[1]{\distributionfunction{thr}_{\ge;#1}}
\newcommand{\thresholddistributionbare}{\distributionfunction{thr}}
\newcommand{\thresholddistributionCDFbare}{\distributionfunction{thr}_{\ge}}

\newcommand{\memorydistribution}[1]{\distributionfunction{mem}_{#1}}

\newcommand{\memorydistributionbare}{\distributionfunction{mem}}

\newcommand{\infdistribution}[1]{\distributionfunction{inf}_{#1}}

\setboolean{twocolswitch}{true}

\begin{document}

\title{\protect
Slightly generalized Generalized Contagion: Unifying simple models of biological and social spreading\\
  \bigskip 
  {\small\textnormal{
      To appear in ``Spreading Dynamics in Social Systems'';\\
      \vspace{-2pt}
      Eds. Sune Lehmann and Yong-Yeol Ahn, Springer Nature.
    }
  }
}

\author{
  \firstname{Peter Sheridan}
  \surname{Dodds}
}

\email{peter.dodds@uvm.edu}

\affiliation{
  Vermont Complex Systems Center,
  Computational Story Lab,
  the Vermont Advanced Computing Core,
  Department of Mathematics \& Statistics,
  The University of Vermont,
  Burlington, VT 05401.
  }

\date{\today}

\begin{abstract}
  \protect
  We motivate and explore the basic features of generalized contagion, a
model mechanism that unifies fundamental models of biological and
social contagion.
Generalized contagion builds on the elementary observation
that spreading and contagion of all kinds involve some form of system memory.
We discuss the three main classes of systems that generalized contagion
affords, resembling:
simple biological contagion;
critical mass contagion of social phenomena;
and an intermediate, and explosive, vanishing critical mass contagion.
We also present a simple
explanation of the global spreading condition
in the context of 
a small seed of infected individuals.
 
\end{abstract}


\maketitle

\section{Introduction}
\label{sec:introduction}

Spreading, construed fully, is everywhere:
the entropically aspirant
diffusive relaxation of all systems;
wave motion, for which ubiquitous is assigned
with no overstatement;
in the propagation of earthquakes;
the expansion of species range, so often involving people;
power blackouts, now able to affect large fractions of the world
population through system growth;
the repeated bane of global pandemics;
economic prosperity and misery;
and 
the talk of the famously talked about.
And understanding how myriad entities spread between people---from
diseases to stories, both true and false---is central
to our scientific understanding of large populations.

Used for good, as the trope goes, a deep knowledge
of contagion mechanisms---contagion science---is necessary
to help in our collective efforts to produce
a world where individuals can flourish.
Used for bad, a path scientific knowledge always offers,
malefactors will be empowered in
the persuasion and manipulation of populations
or the breaking of financial systems.
To prevent negative and catastrophic outcomes,
contagion science should be able to provide
us with algorithms for system defense.

There remain many open questions on contagion.
How many types of spreading and contagion mechanisms are there?
How can we identify and categorize real-world contagions?
But we have only recently moved from the data-scarce period of studying social
phenomena to the start of the data-rich stage,
and contagion science is still very much developing

Our goal in this piece is constrained to
revisiting our 2004 revisiting of basic mathematical models of contagion
surrounding one question~\cite{dodds2004a,dodds2005a}:
Can we connect models of disease-like and social contagion?

We call the process we constructd for this objective `generalized contagion'.
We will give a straightforward explanation of
the model here and discuss its most important features.

An incidental contribution with generalized contagion
was to make memory a primary ingredient.
For contagion, memory comes in many forms,
for example, 
in the development of protection against an infectious disease through an immune response,
or through recalling past exposures to some kind of social influence.
The core models of biological and social contagion
incorporate only the simplest kind of memory, that of the present state.

In proceeding, we first outline the independent
and interdependent interaction models of
biological and social contagion.
Apart from standing as the footing of our generalized model,
we will also preserve certain framings and notations.
We then describe our model of generalized contagion
and discuss the three universality classes of systems identified
in the context of small seeds leading to global spreading.

\section{Independent interaction models of biological contagion}

In mathematical epidemiology,
the standard model~\cite{murray2002a}
was first put forward in the 1920s
by Reed and Frost
and formalized
by Kermack and McKendrick~\cite{kermack1927a,kermack1932a,kermack1933a}.
These models came to be generally referred to as SIR
models in reference to the three epidemiological states:
\begin{itemize}
\item 
  Susceptible;
\item 
  Infective (or Infectious);
\item 
  Recovered
  (or Removed or Refractory).
\end{itemize}
Individuals cycle through the states \textbf{S} to \textbf{I} to \textbf{R}
(and then back to \textbf{S} for an SIRS model).
The behavior of these initial models was described by differential equations
but can be easily realized as a discrete time system,
and we will use the latter
framework for our generalized model.
SIR models are also mass action type models, meaning individuals are represented
as normalized fractions of a population which randomly interact with each other.

To connect notation across different models, we will write
the fractions in the three states as
$S_{t}$,
$\phi_{t}$ (normally $I_{t}$),
and
$R_{t}$.
We must have the constraint $S_{t} + \phi_{t} + R_{t} = 1$.
There is no memory in these systems other than the current balance
of Susceptibles, Infectives, and Recovereds.

\begin{figure}
  \centering
  \includegraphics[width=0.6\columnwidth]{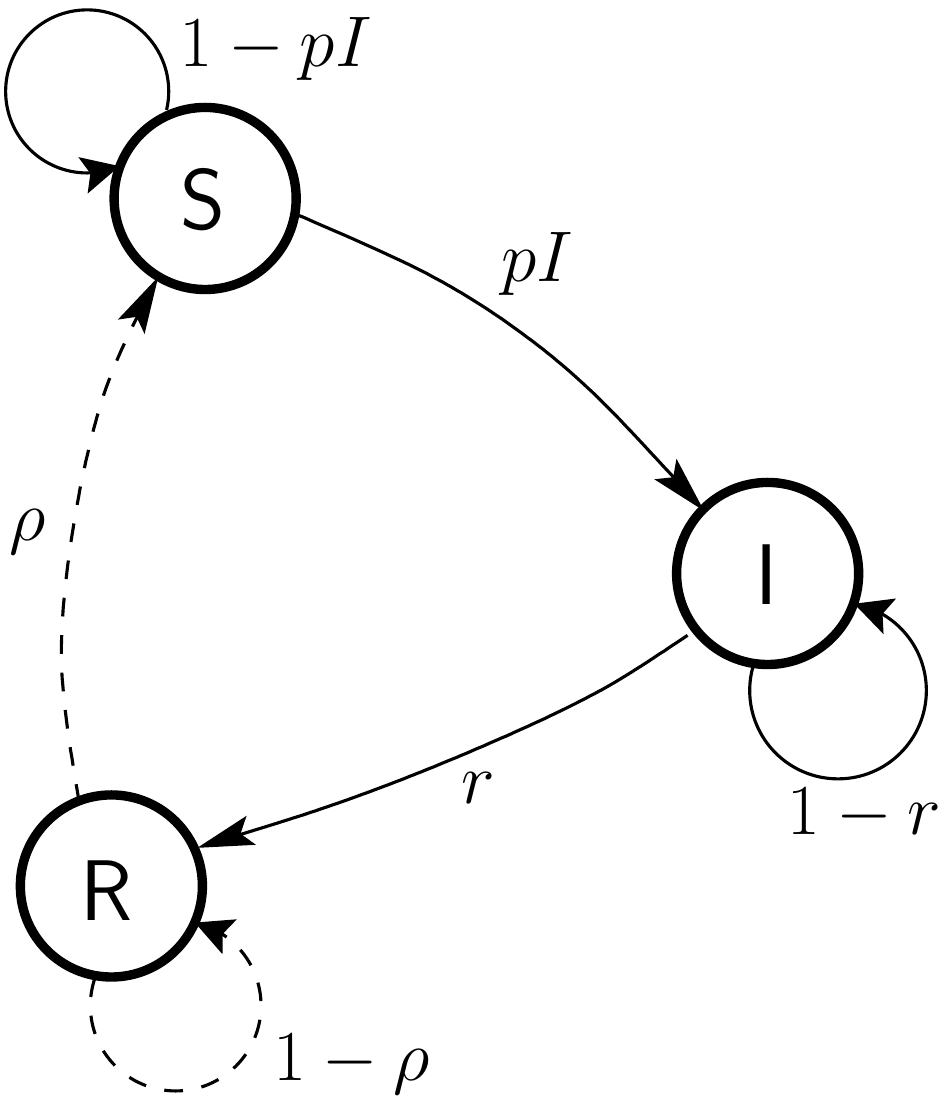}
  \caption{
    Update mechanism for an example discrete version
    of the basic SIR model.
    Individuals may be susceptible (state \textbf{S}),
    infective (state \textbf{I}),
    and recovered (state \textbf{R}).
    The three transition probabilities are
    $\infectionprob$ for being infected given contact with infected
    ($S \rightarrow I$),
    $r$ for recovery
    ($I \rightarrow R$),
    and
    $\rho$ for loss of immunity
    ($R \rightarrow S$).
    The model's complication lies
    in the nonlinear term
    involved in the transition of susceptibles to infectives.
  }
  \label{gcontchap.fig:SIRmodel}
\end{figure}

Fig.~\ref{gcontchap.fig:SIRmodel} shows an example automata
for the independent interaction model when time is discrete.
From the point of view of an individual agent in a
discrete time SIR system, they interact independently,
at each time step connecting with a
Susceptible,
Infective,
or Recovered.
The probabilities of each interaction are equal to the normalized fractions
$S_{t}$, $\phi_{t}$, and $R_{t}$.
When Susceptibles interact with Infectives (occurring with probability $\phi_{t}$),
they themselves become Infective with probability $\infectionprob$.
Regardless of their interactions, Infectives recover with a probability $r$
and Recovereds become Susceptibles with
probability $\rho$ (for SIR models, $\rho=0$, while for SIRS models, $\rho > 0$).

A traditionally key quantity in mathematical
epidemiology is the Reproduction Number $R_0$
[which is terrible notation given we already
have state \textbf{R} and $R_{t}$].
The Reproduction Number is
the expected number of infected individuals resulting
from the introduction of a single initial infective.
The Reproduction Number is easily interpreted
and leads to an Epidemic threshold: If $R_0 > 1$, an `epidemic' occurs.
As with many complex systems, the focus on a single
number as a diagnostic is always fraught, and
the Reproduction Number ultimately combines too many
aspects of the disease itself and population interaction
patterns, rendering it a deceptive measure~\cite{watts2005a}.
Nevertheless, for simple models $R_0$ is important
and the notion of an Epidemic Threshold is more generally essential.

For our simple discrete model, we can compute $R_0$ easily.
We introduce one Infective into a randomly mixing population of Susceptibles.
At time $t=0$, this single Infective randomly bumps into a Susceptible
who is infected with probability $\infectionprob$.
The single Infective remains infected with probability $(1-r)^t$
at time $t$, having attempted to infect $t$ Susceptibles by this point.
The expected number infected by original Infective is therefore:
\begin{gather}
  R_0 = \infectionprob + (1-r)\infectionprob + (1-r)^2\infectionprob + (1-r)^3\infectionprob + \ldots \nonumber \\
  = \infectionprob \frac{1}{1 - (1-r)}
  = \infectionprob/r,
  \label{gcontchap.eq:R0calc}
\end{gather}
and the disease spreads in this system if
\begin{equation}
  R_0 = \infectionprob/r > 1.
  \label{gcontchap.eq:epidemicthreshold}
\end{equation}

\begin{figure}
  \includegraphics[width=\columnwidth]{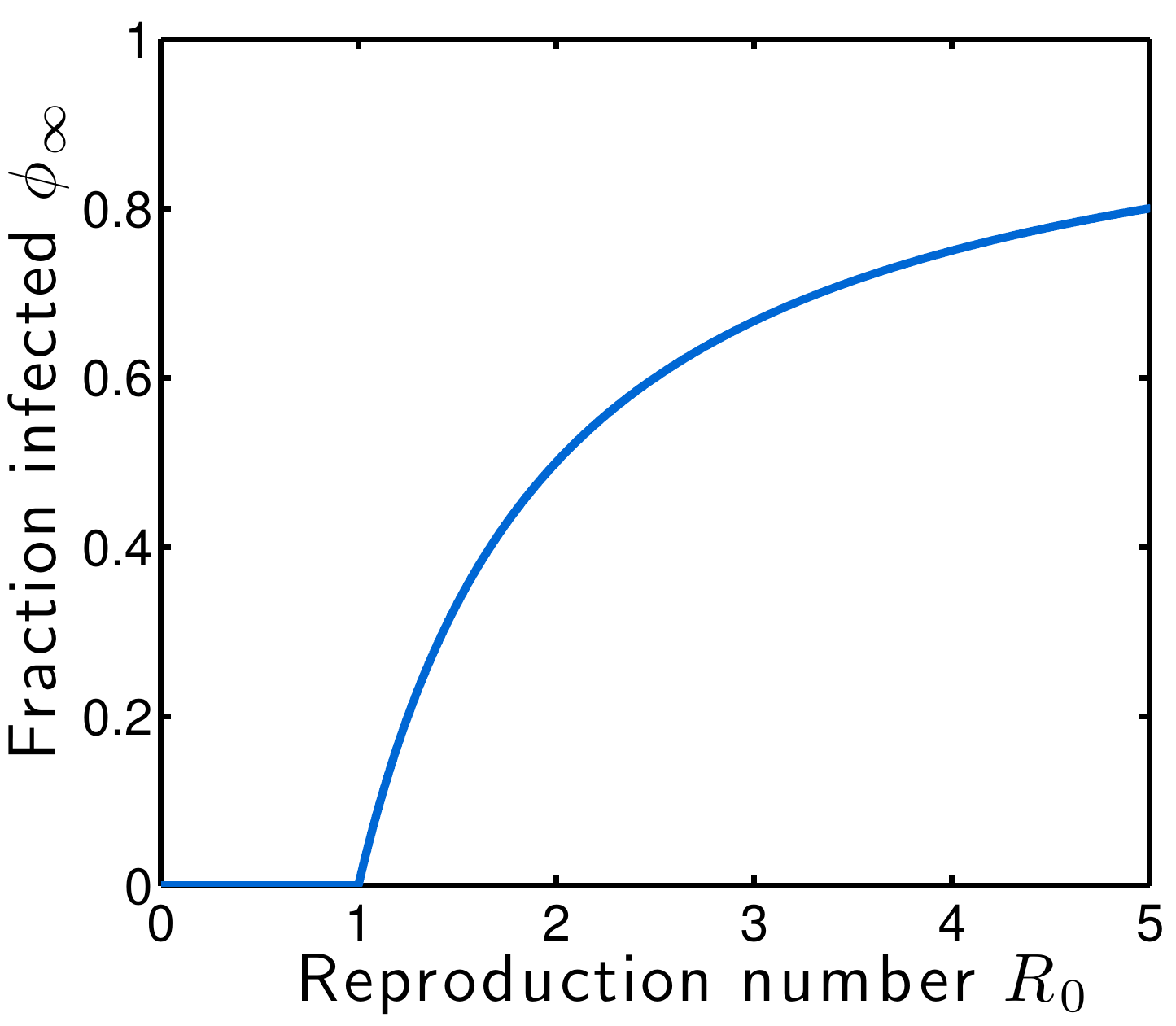}
  \caption{
        Stylized example plot of the final fractional size of a spreading event
    for SIR type models.
    The reproduction number $R_0 = \infectionprob/r$
    (\Req{gcontchap.eq:R0calc})
    acts as a phase parameter
    with a continuous phase transition occurring at $R_0=1$,
    the epidemic threshold.
  }
  \label{gcontchap.fig:epidemicthreshold}
\end{figure}

Fig.~\ref{gcontchap.fig:epidemicthreshold}
shows an example of epidemic threshold from
our elementary SIR model where the tunable
parameter is
the Reproduction Number
$R_0
=
\infectionprob/r.
$
The final fraction infected exhibits
a continuous phase transition
(technically a transcritical bifurcation~\cite{strogatz1994a}).
The epidemic threshold is a powerful story
arising from a simple model.

\section{Interdependent interaction models of social contagion}

\begin{figure}
  \includegraphics[width=\columnwidth]{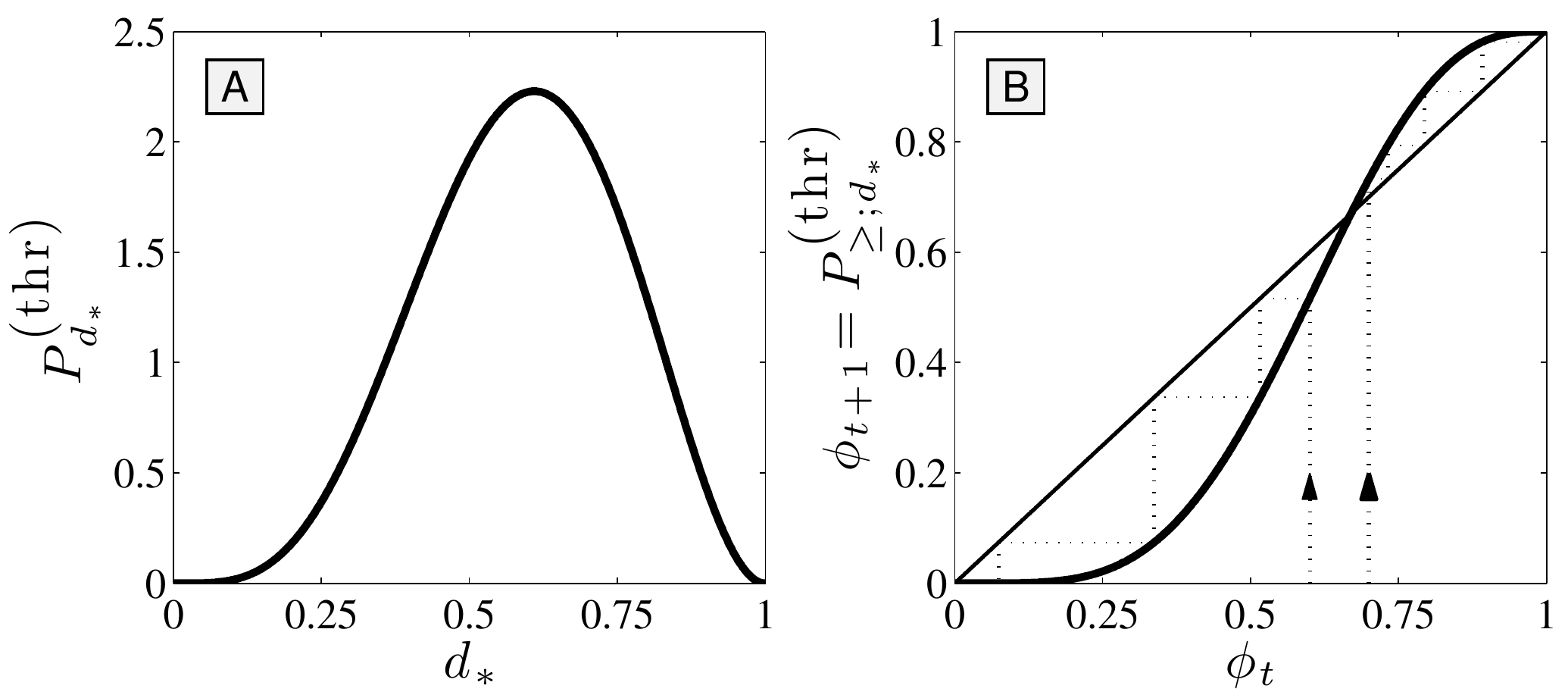}
  \caption{
    Example of Granovetter's model
    reflecting a Critical Mass system.
    \textbf{A:}
    Distribution of individual thresholds:
    $
    \thresholddistribution{\phi}
    =
    \frac{7}{2}
    \phi^{4}
    (1-\phi)^{2}
    (1-\frac{1}{2}\phi)^{2}$.
    \textbf{B:}
    Map of the interval showing
    the evolution of the model
    per~\Req{gcontchap.fig:thresholdmodel}.
    The two cobweb iterates indicate
    how a critical mass is required initially
    for the contagion to be self-sustained and grow.
  }
  \label{gcontchap.fig:thresoldmodel}  
\end{figure}

In spite of the basic SIR model's failings to represent biological
contagion accurately in all cases and particularly at large scales,
it has enjoyed a long tenure.
There have also been overly courageous attempts to use SIR and its sibling models beyond
disease spreading including
the adoption of ideas and beliefs~\cite{goffman1964a},
the spread of rumors~\cite{daley1964a,daley1965a},
the diffusion of innovations~\cite{bass1969a},
and
the spread of fanatical behavior~\cite{castillo-chavez2003a}.

And while some kinds of social contagion may be disease-like, it is
clearly of a different nature for the most part.
One of the major departures is due to the fact that people
take in information from potentially many sources and weigh
their inputs relatively.
This observation gives rise to the notion of
thresholds, first used in modeling in the early 1970s by
Schelling in his efforts to understand segregation~\cite{schelling1971a,schelling1978a}
(the so-called tipping of neighborhoods, and the origin of ``Tipping Point'').
Schelling's model played out (literally) on a chessboard and was manifestly spatial.

Later in the same decade and inspired in part by Schelling's work,
Granovetter produced a distilled mass action threshold model which would become famous
in its own right.
While social contagion is arguably more multifaceted that biological contagion,
Granovetter's model will serve as our elemental model here.

An individual in Granovetter's model may be
framed as having a choice of adopting a behavior
or not based on their perception of that behavior's
popularity.
Each individual $i$ has a threshold
$\dstari \in [0,1]$ drawn from a population-level
threshold distribution $\thresholddistributionbare$ at $t=0$.
We can preserve the SIR model framing
of two states: \textbf{S} and \textbf{I}, with infectives
being those who have adopted the behavior.
We will continue to use $\phi_t$
as the fraction individuals who are infected.

At each time step, if individual $i$ observes
the fraction \textbf{I} of the total population expressing
the behavior as meeting or exceeding
their threshold $\dstari$, then they adopt
the behavior.
The system iterates forward, potentially reaching
an asymptotic state.

Without any spatial structure, all of the interesting dynamics of Granovetter's model
is generated purely by the threshold distribution
$\thresholddistributionbare$.
We are in fact in the realm of maps of the interval,
the territory where so many extraordinary findings
have been made for dynamical systems and chaos~\cite{strogatz1994a}.
The time evolution of Granovetter's model can be written down as:
\begin{equation}
  \phi_{t+1}
  =
  \int_{0}^{\phi_{t}}
  \thresholddistribution{u}
  \dee{u}.
  \label{gcontchap.fig:thresholdmodel}
\end{equation}
The fraction infected in the next time step $\phi_{t+1}$
will be exactly the fraction whose
threshold is exceeded by the current fraction infected $\phi_{t+1}$.

Writing $\thresholddistributionCDFbare$ as the cumulative function of
$\thresholddistributionbare$,
we have, compactly, that
\begin{equation}
  \phi_{t+1}
  =
  \thresholddistributionCDF{\phi_{t}}.
  \label{gcontchap.fig:thresholdmodel_cumulative}
\end{equation}
The dynamics of Granovetter's model are thus inscribed in
$\thresholddistributionCDFbare$
particularly in $\thresholddistributionCDFbare$'s fixed points and relative slopes.
As an example, Fig.~\ref{gcontchap.fig:thresoldmodel} shows how Granovetter's
model may represent a critical mass phenomenon.
Fig.~\ref{gcontchap.fig:thresoldmodel}A gives
the distribution $\thresholddistributionCDFbare$ of individual thresholds showing
a middle tendency.
There are very few extremely gullible people ($\dstar \simeq 0$)
and very difficult to influence ones ($\dstar \simeq 1$).
In Fig.~\ref{gcontchap.fig:thresoldmodel}B,
the cumulative function with some example cobweb iterates~\cite{strogatz1994a}
show that if
the initial fraction infected is above the internal fixed point,
the fraction adopting the behavior rapidly approaches 1,
while any initial fraction starting below the fixed point
will see the behavior die out.
The initial adoption level $\phi_0$ must be generated by
an exogenous mechanism (e.g., education, marketing)
and then the purely imitative dynamics of the system take off.

Granovetter's model and its variants are rich in
dynamics and avenues of analysis~\cite{granovetter1983a,granovetter1986a,granovetter1988a,watts2002a,watts2007a}.
In reintroducing spatial interactions,
Watts transported Granovetter's model to random networks~\cite{watts2002a}
showing that limiting an individual's awareness to a small set
of neighbors on a network could lead to large-scale, potentially catastrophic and
unexpected spreading~\cite{watts2002a}.
And in moving to more structured, socially realistic networks,
even more surprising dynamics open up as possibilities~\cite{watts2009a,dodds2013a,harris2014a}.

\section{Generalized Contagion Model}
\label{sec:model}

The SIR and threshold models are of course intended to be simple,
extracting the most amount of story
from the least amount of stage setting.
But let's list some standard ``I have two comments''-type complaints anyway.
As we have trumpeted, both models involve no memory other than of the current state
traditional disease models assume independence of infectious events.
Threshold models only involve proportions: $17/73 \equiv 170/730$.
Threshold models also ignore the exact sequence of influences
and assume immediate and repeated polling.
Other issues applying to both models, and ones that we will not attend to here,
include the choice between continuous and discrete time,
synchronous updating for discrete time models,
and the dominant assertion of random mixing populations
(even so, network effects are only part of the story as
media provides population-scale and sub-population scale signals).
(Standard random scientist issue: ``You did not cite my work [which you will find out is not related].'')

We would like to bring these basic models of biological and social
contagion together, and, if this is possible,
see if we can gain some new knowledge about contagion processes in general.
Adding memory will be the way forward.
Memory has been successfully incorporated
into other kinds of social contagion models with
a view to modeling real world behavior online~\cite{weng2012a,gleeson2016a}.

We explain generalized contagion in the context
of a random mixing model acting on
a population of $N$ individuals.
We will again have the three states
\textbf{S}, \textbf{I}, and \textbf{R},
for
susceptibles,
infectives,
and
recovereds.

The major variation on the previous models
is that each individual has a fixed memory length
$T$ drawn from
a distribution
$\memorydistributionbare$
with
$1 \le T \le T_{\textnormal{max}}$.
In~\cite{dodds2004a} and
\cite{dodds2005a}, $T$ was the same for all individuals.
At all times, individual $i$ possesses a record of
their last
$T_i$
interactions,
a kind of ticker tape memory.
Each entry in individual $i$'s memory will be either
zero or a dose received from a successful
interaction with an infective (details below).

As for Granovetter's model,
we allow for a general
threshold distribution, $\thresholddistributionbare$.
All nodes randomly select a threshold $\dstar$
using $\thresholddistributionbare$,
and thresholds remain fixed.
Both memory and thresholds could
be made to vary with time though we do not
do this here.

Here's the game play for each step.

At each time step,
regardless of their current state,
each individual $i$ will 
interact with a randomly chosen individual $i'$ from the population.
Next:
\begin{enumerate}
\item 
  Individual $i'$ will be an infective
  with probability $\phi_t$,
  the current fraction of infectives.
  \begin{enumerate}
  \item 
    With probability
    $\infectionprob$,
    a dose is successfully
    transmitted to $i$---an exposure.
    The dose size $d$ will be drawn from a distribution
    $\dosedistributionbare$.
  \item
    With probability
    $1-\infectionprob$,
    $i$ will not be exposed and they
    will record a dose size $d=0$.
  \end{enumerate}
\item
  Individual $i'$ will not be an infective
  with probability $1-\phi_t$ and
  $i$ will record $d=0$ in its memory.
\end{enumerate}

For the SIR model,
$\infectionprob$
was the probability of a successful infection
whereas now it is the probability of a successful
transmission of a dose which is in turn probablistic.

\begin{figure}
  \centering
  \includegraphics[width=0.6\columnwidth]{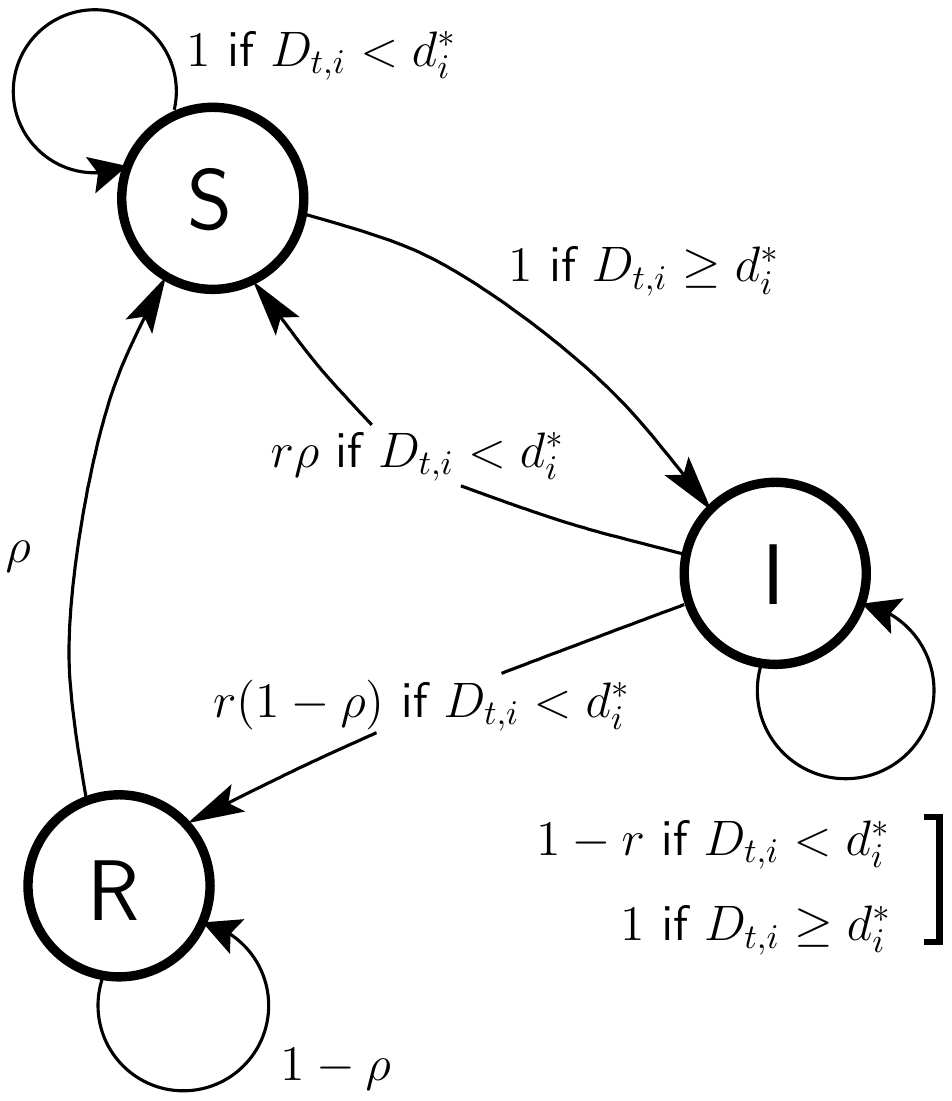}
  \caption{
    Mechanism of the generalized contagion model,
    developing from the same template
    used for the SIR model in
    Fig.~\ref{gcontchap.fig:SIRmodel}.
  }
  \label{gcontchap.fig:transitions}
\end{figure}

Node $i$'s updates its current dosage level
$D_{t,i}$
as the sum
of its last $T_i$ doses:
\begin{equation}
  D_{t,i}
  =
  \sum_{t'=t-T_i+1}^{t}
  d_{t,i}.
  \label{gcontchap.eq:dosage}
\end{equation}

We can now define transition probabilities for
individuals in each of the three states.
As shown in Fig.~\ref{gcontchap.fig:transitions}:
\begin{itemize}
\item 
  \textbf{S} $\Rightarrow$ \textbf{I}:
  Infection occurs if individual $i$'s `threshold' is exceeded:
  \begin{equation}
    D_{t,i}
    \ge
    \dstari.
    \label{gcontchap.eq:Ddtari}
  \end{equation}
\item
  \textbf{I} $\Rightarrow$ \textbf{R}:
  Only if  $D_{t,i} < \dstari $,
  individual $i$ may recover to state
  R
  with probability $r$.
\item
  \textbf{R} $\Rightarrow$ \textbf{S}:
  An individual $i$ may become susceptible again with
  probability $\rho$.
  A detail here is that we allow nodes that arrive
  in state \textbf{R} an immediate chance of returning
  to \textbf{S} in the same time step.
  Nodes in state \textbf{R} are immune and will remain in state \textbf{R} even
  if their dosage level $D_{t,i}$ exceeds
  their threshold.
\end{itemize}

\section{Analysis}
\label{gcontchap.sec:analysis}

We now perform some basic analyses of the generalized contagion
model with a focus on determining the potential for a small seed to
lead to a global spreading event, and
characterizing the abruptness of that
spreading if it is possible.
In doing so, we will show how the dynamics of the SIR and threshold models
are contained within that of generalized contagion.

Expanding on the results of~\cite{dodds2004a,dodds2005a},
the key quantity for our analysis is 
the probability that a randomly selected threshold $\dstar$
will be exceeded by $k$ randomly selected doses
drawn from $\dosedistributionbare$.
Using the notation
$\infdistribution{k}$
we have
\begin{equation}
\infdistribution{k}
=
\int_{0}^{\infty} \dee{\dstar}
\thresholddistribution{\dstar}
\Prob{
 \sum_{j=1}^{k} d_j \ge \dstar
}.
\label{gcontchap.eq:Pk}
\end{equation}
The integral is over all thresholds $\dstar$,
and the probability in the integrand is the
cumulative distribution of the convolution
of $k$ copies of the dose distribution
$\dosedistributionbare$.

The probabilities
$\infdistribution{1}$
and
$\infdistribution{2}$
will prove to be essential.
In particular,
$\infdistribution{1}$,
the probability that one randomly chosen dose will exceed
one randomly chosen threshold
will determine if
SIR-like dynamics are possible.
The quantity
$\infdistribution{1}$
can be interpreted 
as the population fraction of the
most ``vulnerable'' individuals~\cite{watts2002a}.
Whatever the length of memory $T$ of these individuals,
they typically require only one dose to become infected,
and their high susceptibility enables the contagion to spread.
This is a harder story to see and many are readily
taken by the simpler, naive
ones of ``super-spreaders'' and ``influentials.''

We will consider the SIS version,
$\rho=1$,
and
the case of immediate recovery once an individual dosage drops
below its threshold,
$r=1$.
Although more difficult, some analytic work can be carried out if these
probabilities are reduced below 1 (many variations are explored in~\cite{dodds2005a}), and,
of course, simulations
can always be readily performed.

As with many dynamical systems problems,
we are able to determine the main features
of the $\rho=r=1$ generalized contagion system
by examining the system's fixed points
which follow from the system's
update equation:
\begin{equation}
  \phi_{t+1}
  =
  \sum_{T=1}^{T_{\textnormal{max}}}
  \memorydistribution{T}
  \sum_{k=1}^{T}
  \binom{T}{k}
  (\infectionprob\phi_{t})^{k}
  (1-\infectionprob\phi_{t})^{T-k}
  \infdistribution{k}.
  \label{gcontchap.eq:general-updateequation}
\end{equation}
Reading through the right hand side of this fixed point equation,
we first have the probability that a randomly chosen
individual has a memory of length $T$,
$\memorydistribution{T}$.
The inner sum then computes the probability that an
individual with memory of length $T$'s threshold is exceeded after receiving
all possible numbers of positive doses,
$k=1$ to $k=T$.

To find a closed form expression
for the fixed points of the system,
we set $\phi_{t+1}=\phi_{t}$:
\begin{equation}
  \phifix
  =
  \sum_{T=1}^{T_{\textnormal{max}}}
  \memorydistribution{T}
  \sum_{k=1}^{T}
  \binom{T}{k}
  (\infectionprob\phifix)^{k}
  (1-\infectionprob\phifix)^{T-k}
  \infdistribution{k}.
  \label{gcontchap.eq:general-fixedpts}
\end{equation}
In general, curves for $\phifix$
as a function
of the exposure probability
$\infectionprob$
will need to be determined numerically.
However, for the question of whether a small seed
may lead to a global spreading event or not,
we can use
\Req{gcontchap.eq:general-fixedpts}
to find universal results.

Expanding
\Req{gcontchap.eq:general-fixedpts}
for $\phifix$ near 0 we obtain:
\begin{equation}
  \phifix
  =
  \sum_{T=1}^{T_{\textnormal{max}}}
  \memorydistribution{T}
  T
  \infectionprob\phifix
  \infdistribution{1}
  +
  O(\phifix^2).
  \label{gcontchap.eq:general-fixedpts-smallphi}
\end{equation}
Taking $\phifix \rightarrow 0$,
we find the critical exposure probability for the system
is therefore given by
\begin{equation}
  \infectionprobcrit
  =
  \frac{1}
  {
    \tavg{T}
    \infdistribution{1}
  },
  \label{gcontchap.eq:pcrit}
\end{equation}
where
$
\tavg{T}
=
\sum_{T=1}^{T_{\textnormal{max}}}
T
\memorydistribution{T}
$
is the average memory length
(if all individuals have
a memory of uniform length $T_{\ast}$,
as assumed in~\cite{dodds2004a} and~\cite{dodds2005a},
\Req{gcontchap.eq:pcrit}
reduces to
$
\infectionprobcrit
=
1/[T_{\ast}\infdistribution{1}]$.)
We interpret $\infectionprobcrit$
in the same way as the epidemic threshold
of the SIR model.
Global spreading from small seeds will
occur if
$p > \infectionprobcrit$,
and this will only be feasible if
the condition for an epidemic threshold is satisfied:
\begin{equation}
  \infectionprobcrit
  <
  \infectionprob
  < 1.
  \label{gcontchap.eq:epidemicthreshold}
\end{equation}
If instead
$\infectionprobcrit > 1$,
then our system will be more social-like.
As we will show below,
an initial critical mass will be needed
for spreading to take off,
if any spreading is possible at all.

To make the epidemic threshold criterion
for generalized contagion 
intuitive,
we can combine
Eqs.~\ref{gcontchap.eq:pcrit}
and~\ref{gcontchap.eq:epidemicthreshold}
to form the condition:
\begin{equation}
  \left(
  \infectionprob
  \tavg{T}
  \right)
  \cdot
  \infdistribution{1}
  > 1.
  \label{gcontchap.eq:epidemicthreshold2}
\end{equation}
For a small seed to take off,
the interpretation
of~\Req{gcontchap.eq:epidemicthreshold2}
tracks as follows.
Consider one infected individual
at $t=0$ with a one off dose in their memory
exceeding their threshold.
They will randomly interact with
$T$ different uninfected individuals before they
themselves recover.
The expected number of exposures
they will produce in this time
is
$\infectionprob \tavg{T}$,
the first term in
\Req{gcontchap.eq:epidemicthreshold2}.
Because the seed set of infectives
is infinitesimally small,
each susceptible individual interacted with
by an infective will receive at most one dose.
And this dose will infect them with
probability
$\infdistribution{1}$,
the second term in
\Req{gcontchap.eq:epidemicthreshold2}.
Thus,
$
\infectionprob
\tavg{T}
\cdot
\infdistribution{1}
$
is the expected number of new infectives
due to one infective, equivalent
to the reproduction number $R_0$ of the SIR model.
In short,
\Req{gcontchap.eq:epidemicthreshold2}
is the statement that one infective
begets at least one new infective,
leading to an initial exponential growth
of the contagion.

We now need to take some more care as
the epidemic threshold for generalized contagion
is not as simple as that of SIR contagion.
If $\infectionprobcrit < 1$,
we must consider whether the transition
is continuous or discontinuous.
As we saw with the example
in Fig.~\ref{gcontchap.fig:epidemicthreshold},
it is always the former for the SIR model.

\begin{figure*}
  \includegraphics[width=\textwidth]{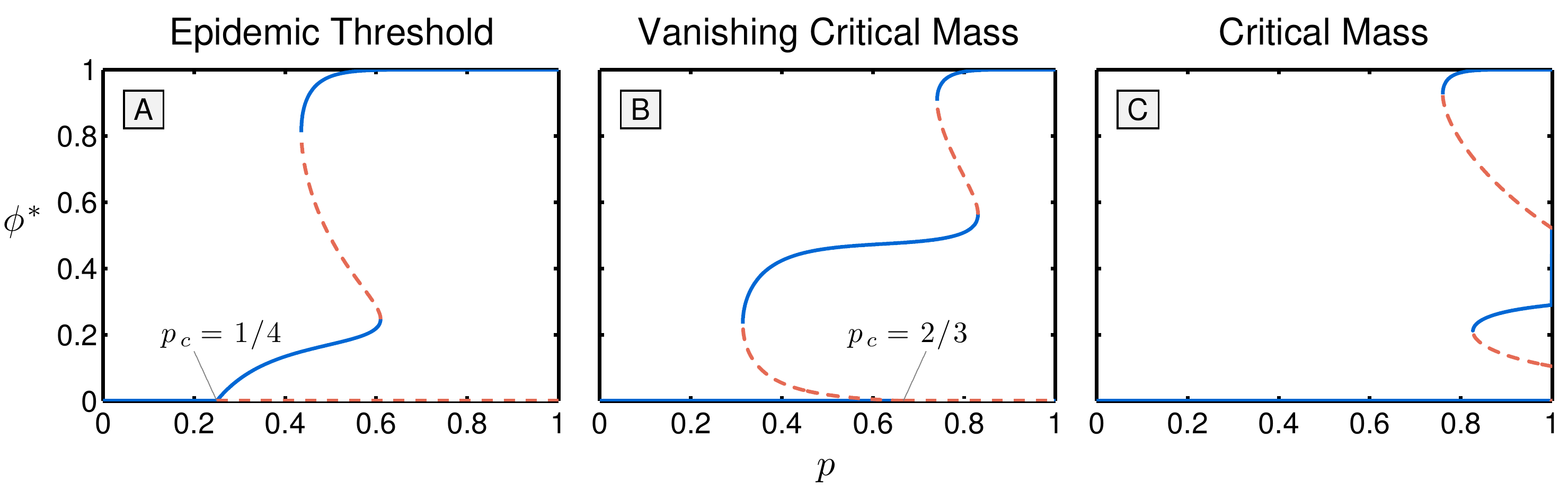}
  \caption{
    Examples of the three main universality classes
    with added bifurcative embellishments, arising
    solely from variable threshold distributions.
    In the epidemic threshold and vanishing
    critical mass cases of
    \textbf{A} and \textbf{B},
    phase transitions for
    $\infectionprob < 1$ are both apparent
    but are strikingly different.
    The continuous phase transition
    in \textbf{A} means the system's behavior
    does not change abruptly as
    $\infectionprob$
    moves above
    $\infectionprobcrit$
    for a small seed $\phi_{0}$.
    The discontinuous phase transition of
    \textbf{B} however means that the growth
    will be sudden and large.
    Vanishing critical mass models with $r=1$
    have
    $
    \infdistribution{2}
    >
    2\infdistribution{1}
    $
    which can be interpreted as meaning
    that the mutual effect of two doses
    is greater than their direct sum would suggest.
    In \textbf{C},
    we see a critical mass
    system for which only a non-zero fraction
    must be initially infected for the contagion
    to maintain and spread.
    Mathematically, $\infectionprobcrit > 1$,
    so no small seed can take off.
    In all three cases, initial seeds will
    grow if the fixed point curve directly
    below them is unstable (and necessarily
    the one above will be stable).
    Simulation details (adapted from~\cite{dodds2005a}):
    $r=\rho=1$,
    $\memorydistribution{d}=\delta_{T,1}$,
    and
    $\dosedistribution{d}=\delta_{d,1}$.
    \textbf{A:}
    $\dosedistribution{\dstar} = 0.2\delta(d-1) + 0.8\delta(d-6)$;
    \textbf{B:}
    $\dosedistribution{\dstar} = 0.075\delta(d-1) + 0.4\delta(d-2) + 0.525\delta(d-12)$;
    and
    \textbf{C:}
    $\dosedistribution{\dstar} = 0.3\delta(d-3) + 0.7\delta(d-12)$.
    All curves were obtained from numerically solving
    \Req{gcontchap.eq:general-fixedpts}.
  }
  \label{gcontchap.fig:threeclasses}
\end{figure*}

If the transition is continuous, then
when
$\infectionprob
=
\infectionprobcrit$
a small
seed at
will not grow,
whereas when the transition is discontinuous,
spreading will take off rapidly.

To test the phase transition's continuity, 
we expand \Req{gcontchap.eq:general-updateequation}
to second order:
\begin{gather}
  \phi_{t+1}
  \simeq
  (\infectionprob\phi_{t})
  \sum_{T=1}^{T_{\textnormal{max}}}
  \memorydistribution{T}
  T
  \infdistribution{1}
  +
  \nonumber \\
  (\infectionprob\phi_{t})^2
  \sum_{T=1}^{T_{\textnormal{max}}}
  \memorydistribution{T}
  T(T-1)
  \left[
    \frac{1}{2}
    \infdistribution{2}
    -
    \infdistribution{1}
    \right]
  \nonumber \\
  =
  (\infectionprob\phi_{t})
  \tavg{T}
  \infdistribution{1}
  +
  \nonumber \\
  (\infectionprob\phi_{t})^2
  \tavg{T(T-1)}
  \left[
    \frac{1}{2}
    \infdistribution{2}
    -
    \infdistribution{1}
    \right].
  \label{gcontchap.eq:general-fixedpts-secondorder}
\end{gather}
Setting
$\infectionprob = \infectionprobcrit$, \Req{gcontchap.eq:pcrit},
we have:
\begin{gather}
  \phi_{t+1}
  \simeq
  \phi_{t}
  +
  (\phi_{t})^2
  \frac{
    \tavg{T(T-1)}
  }{
    \tavg{T}^2
    [\infdistribution{1}]^2
  }
  \left[
    \frac{1}{2}
    \infdistribution{2}
    -
    \infdistribution{1}
    \right].
  \label{gcontchap.eq:general-fixedpts-secondorder2}
\end{gather}
A discontinuous phase transition is
apparent if the fraction infected
$\phi_{t}$
grows
and this evidently occurs if
right hand side of
\Req{gcontchap.eq:general-fixedpts-secondorder2}
is positive, meaning:
\begin{gather}
  \infdistribution{2}
  <
  2\infdistribution{1}:
  \mbox{continuous,}
  \nonumber \\
  \infdistribution{2}
  >
  2\infdistribution{1}:
  \mbox{discontinuous.}
  \label{gcontchap.eq:discontinous-condition}
\end{gather}
We see that the kind of contagion behavior we
observe with social phenomena, that repeated doses
combine superlinearly
$\infdistribution{2}
>
2\infdistribution{1}$,
corresponds with explosive
spreading of a small seed at the critical point.
Discontinuous phase transitions are phase transitions
of surprise---as we increase the exposure probability
$\infectionprob$ starting well below $\infectionprobcrit$,
we see no spreading until 
we reach $\infectionprobcrit$ (or just below depending on $\phi_0$)
when the growth will both be sudden and potentially leading
to a large final fraction of infection.
If repeated doses combine sublinearly,
$\infdistribution{2}
<
2\infdistribution{1}$,
then the final fraction of infections will grow
continuously from 0 as we move past
$\infectionprobcrit$.
Now, this is for the special case of a pure SIS model
and as we later note, the criterion
for a vanishing critical mass model,
$\infdistribution{2}
>
2\infdistribution{1}$,
does not remain so simple as we move
to more complicated models.
So, while we can observe that a sufficiently nonlinear
interaction in doses leads to non-epidemic threshold
model,
we arguably should not have been able to intuit
the simple inequality
$\infdistribution{2}
>
2\infdistribution{1}$
as being the salient test.

We can now assert
that the generalized contagion model produces
three distinct universality classes with
respect to spreading behavior from a small seed.
These are:
\begin{itemize}
\item
  \textbf{Epidemic Threshold Class:}\\
  Criteria:
  \begin{enumerate}
  \item 
  $p_c = 1/(\tavg{T}\infdistribution{1}) < 1$.
  \item 
  $\infdistribution{1} > \infdistribution{2}/2$.
  \end{enumerate}
\item
  \textbf{Vanishing Critical Mass:}\\
  Criteria:
  \begin{enumerate}
  \item 
  $p_c = 1/(\tavg{T}\infdistribution{1}) < 1$.
  \item 
  $\infdistribution{1} < \infdistribution{2}/2$.
  \end{enumerate}
\item
  \textbf{Pure Critical Mass:}\\
  Criteria:
  \begin{enumerate}
  \item
    $p_c = 1/(\tavg{T}\infdistribution{1})$. 
  \item
    \Req{gcontchap.eq:general-fixedpts} is solvable
    with solutions $\phifix(p) \in [0,1]$.
  \end{enumerate}
\end{itemize}

In Fig.~\ref{gcontchap.fig:threeclasses},
we show results from numerically solving
\Req{gcontchap.eq:general-fixedpts}
for three example dose distributions and $T=20$ set uniformly,
(see caption for details; adapted from Fig.~9 in~\cite{dodds2005a}).

The three panels
correspond in order to the three universality
classes.
We emphasize that the universality classes
we find here relate to the kind of critical point
present in the system for $\phifix=0$, if
such a critical point exists.
The details of these systems are unimportant
as many threshold and dose distributions give same
$\infdistribution{k}$.
All solid blue curves indicate stable fixed points
and dashed red curves unstable fixed points.

For the epidemic threshold
in Fig~\ref{gcontchap.fig:threeclasses}A,
we see a continuous
phase transition occurring at $\infectionprobcrit = 1/4$.
Small seeds for $\infectionprob$ above $\infectionprobcrit$
will grow but be constrained.

In Fig.~\ref{gcontchap.fig:threeclasses}B,
the Vanishing Critical Mass class also shows
a epidemic threshold but now the phase
transition is discontinuous.
Tuning the system from below to above
$\infectionprobcrit = 2/3$,
a small seed moves from ineffectual
to suddenly producing successful
global spreading to, roughly, half
of the population.

The fixed point curves for the Critical Mass
model in
Fig.~\ref{gcontchap.fig:threeclasses}C
show the resilience of this third class to
small seeds initiating spreading events.
Only if the initial seed is above the dashed
red curves of unstable fixed points, will
the final extent of spreading be non-zero
(this statement is true for all three classes).

For uniform memory length $T_{\ast}$,
the full linearization near $\infectionprob$
has the form~\cite{dodds2005a}:
\begin{equation}
  \label{gcontchap.eq:transIcalc3}
  \phifix 
  \simeq
  \frac{C_1}{C_2 \infectionprob^2}
  (p-\infectionprobcrit)
  =
  \frac{T_{\ast}^2 P_1^3}
       {(T_{\ast}-1)(P_1 - P_2/2)}
       (p-\infectionprobcrit),
\end{equation}
where from the denominator
we can again see
that
$P_1 - P_2/2 =  0$
locates the transition between
Epidemic Threshold models
and
Vanishing Critical Mass models.

Moving away from systems behavior for small seeds,
in all three examples, we see that the threshold distributions are of
enough variability to produce non-trivial fixed point
curves.
Further, both the Epidemic Threshold and Vanishing Critical
Mass examples also show that
hysteresis dynamics 
(with respect to $\infectionprob$)
are available for Generalized Contagion systems.

If we relax the recovery probability $r$ below
1 and/or elevate the immune state
transition probability $\rho$ above 0,
then we see the same three universality classes
will still emerge.
The conditions for the three classes
will become more complex~\cite{dodds2005a}.
The appealing form of the test separating
Epidemic Threshold and Vanishing Critical Mass models,
$\infdistribution{2}
<
2\infdistribution{1}$,
will no longer be quite so simple.
Analytic results are possible for $r<1$
and $\rho=0$~\cite{dodds2005a} while
systems with $\rho>0$ have not yielded, at
least to our knowledge, to exact treatments.

\section{Concluding remarks}
\label{sec:concludingremarks}

We developed generalized contagion
to demonstrate that a single mechanism could
be shown to produce both disease-like and social-like
spreading behavior.
The observation that memory is a natural aspect
of real-world spreading phenomena proved
to be the binding agent.

The three universal classes of contagion processes
pertain to the spectrum of random-mixing models
and their dynamics in the fundamental initial
condition of an infinitesimally small seed.
We see that dramatic changes in behavior are possible,
particularly in the Vanishing Critical Mass class.

Generalized contagion is also another example of a model where
the vulnerable or gullible population may be more important than
a small group of super-spreaders or influentials~\cite{watts2007a}.

Two avenues for changing dynamics are clear.
One would be to change the model itself through
adjusting its parameters: memory, recovery rates, and the
fraction of individuals vulnerable to 1 or 2 doses.
($T$, $r$, $\rho$,
$\infdistribution{1}$,
and
$\infdistribution{2}$).
Given a  model with fixed parameters,
changing the system's behavior would be possible by changing
the probability
of exposure ($\infectionprob$)
and/or
the initial fraction infected ($\phi_0$).

We hope that this overview of generalized contagion serves
as both an introduction to the model itself and
an inspiration for the many possible adjacent
areas in contagion dynamics available for development.
Generalized contagion on social-like networks more
complicated than random networks would be one such
path. While perhaps this work would be resilient
to simple analysis, simulations could prove illuminating.

\clearpage

\end{document}